\documentclass[12pt]{article}
\usepackage{epsfig,amssymb}
\usepackage{color}
\usepackage{graphicx,color}
\usepackage{epstopdf}
\usepackage{float}

\def\appendix{{\newpage\section*{Appendix}}\let\appendix\section%
        {\setcounter{section}{0}
        \gdef\thesection{\Alph{section}}}\section}

\newcommand{\be}{\begin{equation}}
\newcommand{\ee}{\end{equation}}
\newcommand{\bear}{\begin{eqnarray}}
\newcommand{\eear}{\end{eqnarray}}
\newcommand{\ba}{\begin{array}}
\newcommand{\ea}{\end{array}}

\begin{document}

\begin{flushright}
{\tt hep-th/yymmnnn}
\end{flushright}
\vspace{5mm}

\begin{center}
{{\Large \bf The Extended Thermodynamic Properties\\ of
Taub-NUT/Bolt-AdS spaces}\\[14mm]
{Chong Oh Lee}\\[2.5mm]
{\it Department of Physics, Kunsan National University,\\
Kunsan 573-701, Korea}\\
{\tt cohlee@kunsan.ac.kr}}
\end{center}

\vspace{10mm}

\begin{abstract}
We investigate the extended thermodynamic properties of
higher-dimensional Taub-NUT/Bolt-AdS spaces where a cosmological
constant is treated as a pressure. We find a general form for
thermodynamic volumes of Taub-NUT/Bolt-AdS black holes for arbitrary
dimensions. Interestingly, it is found that the Taub-NUT-AdS metric
has a thermodynamically stable range when the total number of
dimensions is a multiple of 4 (4, 8, 12, \ldots). We also explore
their phase structure and find the first order phase transition
holds for higher-dimensional cases.
\end{abstract}
\newpage

\setcounter{equation}{0}
\section{Introduction}

Thermodynamic properties of black holes have been studied for a long
time since it was first noticed that the area of the event horizon
of a black hole is proportional to its physical entropy \cite{Bekenstein:1973ur}.
It was found that their physical quantities are expressed in terms of the
thermal energy $U$, temperature $T$, and entropy $S$ and the first
law of black hole thermodynamics has the similar forms to the first
law of standard thermodynamics \cite{Bardeen:1973gs}.

It has been also studied  in
complete analogy with standard thermodynamic systems. For example,
it was found that there is a phase transition in the
Schwarzschild-AdS black hole \cite{Hawking:1982dh}. Since then, the phase transitions
and critical phenomena in a variety of black hole solutions have
been studied \cite{Lousto:1994jd}-\cite{Myung:2012xc}. In particular, they have been used for
investigation to various thermodynamic issues including higher
dimensional black hole with negative cosmological constant $\Lambda$
\cite{Chamblin:1999tk}-\cite{Banerjee:2011au}.

As widely known, thermodynamic quantities in black hole physics, the
black hole mass $M$, surface gravity $\kappa$ and the horizon area $A$
correspond to the thermodynamic quantities of a physical
system, thermal energy $U$, temperature $T$, and entropy $S$ respectively. However,
comparing the first law of standard thermodynamics with the pressure
and its conjugate to the first law of black hole thermodynamics in
parallel, their counterparts in black hole physics are not quite
captured. Recently, there have been new and interesting developments
in the question, meaning that the cosmological constant $\Lambda$
can be treated as the thermodynamic pressure $p$
\bear\label{pressure}
p=-\frac{1}{8\pi}\Lambda=\frac{u(2u+1)}{8\pi l^2},
\eear
in units where $G=c=\hbar=k_B=1$, and the total number
of dimensions $(d + 1) = 2u + 2$ is even for some integer $u$.
In addition, the black hole mass $M$ is defined as the enthalpy rather than the thermal energy $U$.
This framework is quite natural since when the cosmological constant
is considered as pressure, its conjugate quantity becomes dimensions
of volume. A lot of investigations have been performed in this direction
\cite{Kastor:2009wy}-\cite{Dolan:2014vba}.

Very recently this issue was generalized to Taub-NUT/Bolt-AdS spaces
in \cite{Johnson:2014yja, Johnson:2014pwa} and to Kerr-Bolt-AdS
spaces in \cite{MacDonald:2014zaa}. Interestingly, they found the
thermodynamic volume in Taub-NUT-AdS metric can be negative. In the
context of enthalpy, the positive thermodynamic volume may be
understood as applying the work on the environment (universe) by the
system (the whole black hole) considering the process of forming the
black hole. In addition, the negative thermodynamic volume may be
understood in that the environment (universe) applies work to the
system (Taub-NUT-AdS black hole) in the process of the Taub-NUT-AdS
black hole formation \cite{Johnson:2014yja}.
They also found that there is the first order phase transition from Taub-NUT-AdS to Taub-Bolt-AdS
through exploring the phase structure of a NUT solution and a Bolt solution \cite{Johnson:2014pwa}.

However, their works on the thermodynamic properties of black hole
of the Taub-NUT-AdS metric have only shown the progress in the
four-dimensional cases \cite{Johnson:2014yja, MacDonald:2014zaa,
Johnson:2014pwa}. Furthermore it is well known that their
thermodynamic properties are depending on odd $u$ (the total number
of dimensions: 4, 8, 12, $\ldots$) or even $u$ (the total number of
dimensions: 6, 10, 14, $\ldots$). Here we will investigate
extensively thermodynamic properties in the higher-dimensional
Taub-NUT/Bolt-AdS spaces. It will particularly show that the
Taub-NUT-AdS solution has a thermodynamically stable range as a
function of the temperature for any odd $u$. We also will
demonstrate that there is the transition from Taub-NUT-AdS to
Taub-Bolt-AdS for all odd $u$ only.

The paper is organized as follows: In the next section we
investigate thermodynamic properties in Taub-NUT/Bolt-AdS spaces for
any $u$. We find that thermodynamic quantities such as the entropy,
the enthalpy, the specific heat, the temperature, and the
thermodynamic volume. We discuss their phase structure and their
instability. In the last section we give our conclusion. In Appendix
A it is shown to be explicit forms for their thermodynamic
quantities.

\section{Taub-NUT/Bolt-AdS  black hole}
We start by considering the $(d + 1)$-dimensional-Taub-NUT with negative cosmological constant.
The general solution in the Euclidean section, for a $U(1)$
fibration over a series of the space ${\cal M}^2$ as the base space
$\bigotimes_{i=1}^{u}{\cal M}^2$, is given by
\cite{Chamblin:1998pz}-\cite{Astefanesei:2004kn}
(for the generalized versions of the issue, see e.g., \cite{Mann:2003zh,Mann:2005ra})
\bear
ds^2&=&f(r)\left\{dt_{E}+2N\sum_{i=1}^{u}\cos(\theta_i)d\phi_i\right\}^2
\nonumber\\&+&\frac{dr^2}{f(r)} +(r^2-N^2)\sum_{i=1}^{u}
\Biggr\{d\theta_i^2+\sin^2(\theta_i)d\phi_i^2\Biggr\},
\eear
with the metric function $f(r)$
\bear
f(r)&=&\frac{r}{(r^2-N^2)^u}\int^{r}
\left\{\frac{(a^2-N^2)^u}{a^2}\right.
\nonumber\\&+&\left.\frac{(2u+1)(a^2-N^2)^{u+1}}{l^2 a^2}
\right\}da-\frac{2mr}{(r^2-N^2)^u}.
\eear
Here, $N$ represents a NUT charge for the Euclidean section,
$l$ is a cosmological parameter, $m$ is a geometric mass.
The NUT solution occurs when solving $f(r)|_{r=N}$=0. The inverse of the temperature $\beta$
is obtained by requiring regularity in the Euclidean time
$t_{E}$ and radial coordinate $r$
\cite{Chamblin:1998pz}-\cite{Astefanesei:2004kn}
\bear\label{HT0}
\beta=\left.\frac{4\pi}{f'(r)}\right|_{r=N}=
4(u+1)\pi N,
\eear
where $\beta$  is the period of $t_{E}$.

Using counter term subtraction method, we get the regularized action
\cite{Chamblin:1998pz}-\cite{Astefanesei:2004kn}
\bear\label{ac0}
I_{\rm NUT}=\frac{(4\pi)^{u}N^{2u-1}
(2uN^2-l^2)}{16\pi^{\frac{3}{2}}l^2}
\hspace{3mm}\Gamma(\frac{1}{2}-u)\Gamma(u+1)\beta,
\eear
where the gamma function $\Gamma(t)$ is defined as
\bear
\Gamma(t)=\int_{0}^{\infty}x^{t-1}e^{-x}dx.
\eear
Solving (\ref{HT0}) for the mass at $r=N$, the NUT mass is given as
\bear
m_{\rm NUT}=\frac{N^{2u-1}
\bigg\{l^2-2(u+1)N^2\bigg\}}{\sqrt{\pi}(2u-1)l^2}
\Gamma(\frac{3}{2}-u)\Gamma\left(u+1\right),
\eear
and the entropy is found to be
\cite{Chamblin:1998pz}-\cite{Astefanesei:2004kn}
\bear\label{s0}
S_{\rm NUT}
=\frac{(4\pi)^{u}N^{2u-1}\biggr\{2u(2u+1)N^2-(2u-1)l^2\biggr\}}{16\pi^{\frac{3}{2}}l^2}
\Gamma(\frac{1}{2}-u)\Gamma(u+1)\beta,
\eear
by the Gibbs-Duhem relation $S=\beta M - I$
where $M$ is the conserved mass
\bear\label{conmass}
M=u(4\pi)^{u-1}m.
\eear
As mentioned in the previous section,
its conjugate variable has dimension of volume
when the negative cosmological constant $\Lambda$ is treated
as thermodynamic pressure $p$ (\ref{pressure}).
In the context of the thermodynamic process, we should think of the energy required to create the system
$pV$ as well as the energy of the system $U$ i.e., whole forming the system, which is quite captured by the
enthalpy. Thus, the conserved mass $M$ (\ref{conmass}) is identified with enthalpy $H$
rather than internal energy $U$
\bear
M\equiv H= U+pV,
\eear
which leads to
\bear
H_{\rm NUT}&=&u(4\pi)^{u-1}
\left\{\frac{N^{2u-1}\Gamma\left(\frac{3}{2}-u\right)
\Gamma\left(u+1\right)}{\sqrt{\pi}(2u-1)}\right.\\\nonumber
&&\left.-\frac{16\sqrt{\pi}(u+1)N^{2u+1}\Gamma\left(\frac{3}{2}-u\right)
\Gamma\left(u+1\right)}{u(2u-1)(2u+1)}p\right\}.
\eear
To check whether we perform the computation correctly, we can reproduce the temperature through
\bear
T_{\rm NUT}&=&\left.\frac{\partial H}{\partial S}\right|_p\\
&=& u(4\pi)^{u-1}\left.\frac{\partial N}{\partial S}\right|_p
\left\{\frac{N^{2u-2}\Gamma\left(\frac{3}{2}-u\right)\Gamma\left(u+1\right)}{\sqrt{\pi}}\right.\\\nonumber
&&\left.-\frac{16\sqrt{\pi}(u+1)N^{2u}\Gamma\left(\frac{3}{2}-u\right)\Gamma\left(u+1\right)}
{u(2u-1)}p\right\}\\
&=&\frac{1}{4(u+1)\pi N},
\eear
since, from Eq. (\ref{s0}):
\bear
\left.\frac{\partial S}{\partial N}\right|_p&=&u(u+1)(4\pi)^u N
\left\{\frac{N^{2u-2}\Gamma\left(\frac{3}{2}-u\right)\Gamma\left(u+1\right)}{\sqrt{\pi}}\right.\\\nonumber
&&\left.-\frac{16\sqrt{\pi}(u+1)N^{2u}\Gamma\left(\frac{3}{2}-u\right)\Gamma\left(u+1\right)}
{u(2u-1)}p\right\}^{-1}.
\eear
One the other hand, employing thermal relation $C=-\beta \partial_{\beta}S$, the specific heat is given as
\bear
C_{\rm NUT}
=\frac{u(4\pi)^{u-1}N^{2u-1}\bigg\{2(u+1)(u+1)N^2-(2u-1)l^2\bigg\}}{2\sqrt{\pi}l^2}
\Gamma(\frac{1}{2}-u)\Gamma(u+1)\beta.
\eear
The thermodynamic volume is also obtained as
\bear
V_{\rm NUT}&=&\left.\frac{\partial H}{\partial p}\right|_S\\\nonumber
&=&2(-2\sqrt{\pi})^{2u-1}(u+1)N^{2u+1}\Gamma(-\frac{1}{2}-u)\Gamma(u+1)\\\nonumber
&&+\left.\frac{\partial N}{\partial p}\right|_S
\left\{\frac{u(2u-1)(2u+1)(2\sqrt{\pi}N)^{2u-2}\Gamma(-\frac{1}{2}-u)
\Gamma(u+1)}{4\sqrt{\pi}}\right.\\\nonumber
&&~~~~~~~~~~~~-\left.\frac{\sqrt{\pi}u(2u-1)(2u+1)(2\sqrt{\pi}N)^{2u-2}\Gamma(-\frac{1}{2}-u)
\Gamma(u+1)}{4}p\right\}\\\label{NUTVol}
&=&-\frac{u(4\pi)^uN^{2u+1}}{2\sqrt{\pi}}\Gamma(-\frac{1}{2}-u)\Gamma(u).
\eear
From this result (\ref{NUTVol}), we can read the thermodynamic volumes of the NUT solution
are negative for all odd $u$
since $\Gamma(-\frac{1}{2}-u)$ produces a positive sign for all odd $u$ only,
and $\Gamma(u)$ does a positive sign for any $u$ due to the property of Gamma function
(Table 1 in Appendix A shows their explicit volume forms for four ($u=1$) to ten ($u=4$) dimensions).\\
Also, due to dimensional scaling arguments, the generalized Smarr formula is given as
\bear\label{smarr}
\frac{1}{2}H-\frac{u}{2u-1}TS+\frac{1}{2u-1}pV=0,
\eear
which is precisely matched with that of static $d$-dimensional black holes with negative cosmological constant
\cite{Caldarelli:1999xj,Kastor:2009wy,Cvetic:2010jb}.\\
Finally, the internal energy of Taub-NUT is obtained as
\bear
U_{\rm NUT}=\frac{u(2u+1)}{8\pi}\left\{(2\sqrt{\pi}N)^{2u-1}
-\frac{(2u+1)(2\sqrt{\pi}N)^{2u-1}N^2}{l^2}\right\}\Gamma(-\frac{1}{2}-u)\Gamma(u+1)
\eear
by an thermal relation $U=H-pV$.\\

\begin{figure}[!htbp]
\begin{center}
{\includegraphics[width=9cm]{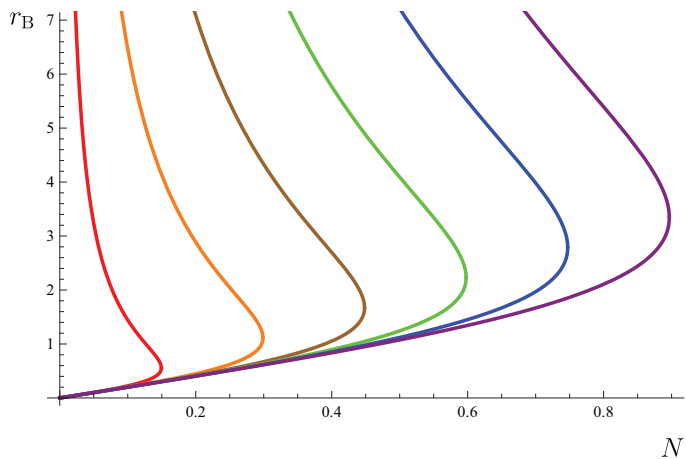}}
\end{center}
\caption{{\footnotesize Plot of the available bolt radii $r_{\rm B}$ as a function of $N$ for $u$
(from left to right) 1 (the dimension of spacetime $2u+2=4$) to 6  (the dimension of spacetime $2u+2=14$)
for cosmological parameter $l=1$.}}
\label{figI}
\end{figure}

Requiring $f(r)|_{r=r_{\rm B}>N}$ and $f'(r)|_{r=r_{\rm M}}=\frac{1}{N(u+1)}$, the Bolt solution
occurs. In Taub-Bolt-AdS metric, the inverse of the temperature, the action,
and the mass are respectively
\bear\label{inverT}
\beta=\left.\frac{4\pi}{f'(r)}\right|_{r=r_{\rm B}}
=\frac{4\pi l^2 r_{\rm B}}{l^2+(2u+1)(r_{\rm B}-N^2)},
\eear
\bear
I_{\rm Bolt}&=&\frac{(4\pi)^{u-1}}{4 l^2}\left[\frac{(2u+1)(-1)^u N^{2u+2}}{r_{\rm B}}
\right.\\\nonumber
&&+\sum_{i=0}^{u}
\left(\begin{array}{l}
u\\
i
\end{array}\right)
\left.(-1)^i N^{2i} r_{\rm B}^{2u-2i}\left\{\frac{l^2}{(2u-2i-1)r_{\rm B}}
-\frac{(2u+1)(u-2i+1)r_{\rm B}}{(2u-2i+1)(u-i+1)}\right\}\right]\beta,
\eear
\bear
m_{\rm Bolt}&=&\frac{1}{2l^2}\left\{\sum_{i=0}^{u}
\left(\begin{array}{l}
u\\
i
\end{array}\right)
\frac{(-1)^i N^{2i}l^2r_{\rm B}^{2u-2i-1}}{(2u-2i-1)}\right.\\\nonumber
&&~~~~~~~~~~~~+\left.(2u+1)\sum_{i=0}^{u+1}
\left(\begin{array}{l}
u+1\\
~~~i
\end{array}
\right)
\frac{(-1)^iN^{2i}r_{\rm B}^{2u-2i+1}}{2u-2i+1} \right\}.
\eear
Here the bolt radii $r_{\rm B}$ is
\bear\label{rB}
r_{\rm B}=\frac{l^2\pm\sqrt{l^4+(2u+1)(2u+2)^2N^2[(2u+1)N^2-l^2]}}
{(2u+1)(2u+2)N},
\eear
and its reality requirement implies
\bear
N\leq \frac{l}{\sqrt{2(u+1)(2u+1)[u+1+\sqrt{u(u+2)}]}}=N_{\rm max},
\eear
where $N_{\rm max}$ denotes  the maximum magnitude of the NUT charge.
As shown in Fig.1, we plot the available bolt radius for the various dimension of spacetime.
It is shown that the $N_{\rm max}$ grows up as the dimension
of spacetime increases. Thus, the magnitude of radii of small bolts in $r_{\rm B}$ (\ref{rB})
increases in the case of the higher dimension.

Using parallel way as in the case of the NUT solution,
the enthalpy $H_{\rm Bolt}$, the entropy $S_{\rm Bolt}$, and thermodynamics volume $V_{\rm Bolt}$
for the Bolt solution yields respectively
\bear
H_{\rm Bolt}&=&\frac{u(4 \pi)^{u-1}}{2}
\left\{\sum_{i=0}^{u}
\left(\begin{array}{l}
u\\
i
\end{array}\right)
\frac{(-1)^i N^{2i}r_{\rm B}^{2u-2i-1}}{(2u-2i-1)}\right.\\\nonumber
&&~~~~~~~~~~~~~~~\left.+\frac{8\pi}{u}\sum_{i=0}^{u+1}
\left(\begin{array}{l}
u+1\\
~~~i
\end{array}
\right)
\frac{(-1)^iN^{2i}r_{\rm B}^{2u-2i+1}}{(2u-2i+1)}p\right\},
\eear
\bear
S_{\rm Bolt}&=&\frac{(4\pi)^{u-1}}{4}
\left[\sum_{i=0}^{u}
\left(\begin{array}{l}
u\\
i
\end{array}\right)
\frac{(2u-1)(-1)^iN^{2i}r_{\rm B}^{2u-2i-1}}{2u-2i-1}
\right.\\\nonumber
&&+\left.\left\{
\sum_{i=0}^{u}
\left(\begin{array}{l}
u\\
i
\end{array}\right)
\frac{8\pi(2u^2+3u-2i+1)(-1)^{i}N^{2i}r_{\rm B}^{2u-2i+1}}{u(u-i+1)(2u-2i+1)}
+\frac{8\pi(2u-1)N^{2u+2}}{ur_{\rm B}}
\right\}p\right]\beta,
\eear

\bear
V_{\rm Bolt}&=&\frac{\pi^{u}(2r_{\rm B})^{u-1}}{2u+1}
\Bigg\{-2(N^2-r_{\rm B}^2)
\left(1-\frac{N^2}{r_{\rm B}^2}\right)^u\\\nonumber
&&~~~~~~~~~~~~~~~~~~~+r_{\rm B}^2\left(\frac{N}{r_{\rm B}}\right)^{2u+1}
{\rm B} \left(\frac{N^2}{r_{\rm B}^2},\frac{1}{2}-u,u+1\right)\Bigg\},
\eear
where the incomplete beta function
${\rm B} \left(x,a,b\right)$ is
defined as
\bear
{\rm B}\left(x,a,b\right)=\int_{0}^{x} t^{a-1}(1-t)^{b-1}dt.
\eear
Employing the thermal relation $U=H-pV$, the internal energy of AdS-Taub-Bolt $U$ is given as
\bear
U_{\rm Bolt}=\frac{(4\pi)^{u-1}}{4l^2}uN^{2u-1}\Bigg\{(2u+1)N^2-l^2\Bigg\}
{\rm B} \left(\frac{N^2}{r_{\rm B}^2},\frac{1}{2}-u,u+1\right).
\eear

Let us start with the action difference of Taub-NUT-AdS and Taub-Bolt-AdS
in order to study the phase structure of the Taub-NUT/Bolt-AdS system.
Their action difference, ${\cal I}_{D}$ is defined as
\bear\label{diffaction}
I_{\rm Bolt}-I_{\rm NUT}\equiv {\cal I}_{D} &=& \frac{(4\pi)^u}{8r_{\rm B}l^2}\left[2N(N^2-r_{\rm B}^2)
(u+1)\left(1-\frac{N^2}{r_{\rm B}^2}\right)^u\right.\\\nonumber
&&+(2uN^2-l^2)\left\{(u+1)N^{2u}{\rm B}
\left(\frac{N^2}{r_{\rm B}^2},\frac{1}{2}-u,u+1\right)\right.\\\nonumber
&&~~~~~~~~~~~~~~~~~~~~~
\left.\left.-\frac{2N^{2u}}{\sqrt{\pi}}\Gamma(\frac{1}{2}-u)\Gamma(2+u)\right\}\right].
\eear

\begin{figure}[!htbp]
\begin{center}
{\includegraphics[width=8cm]{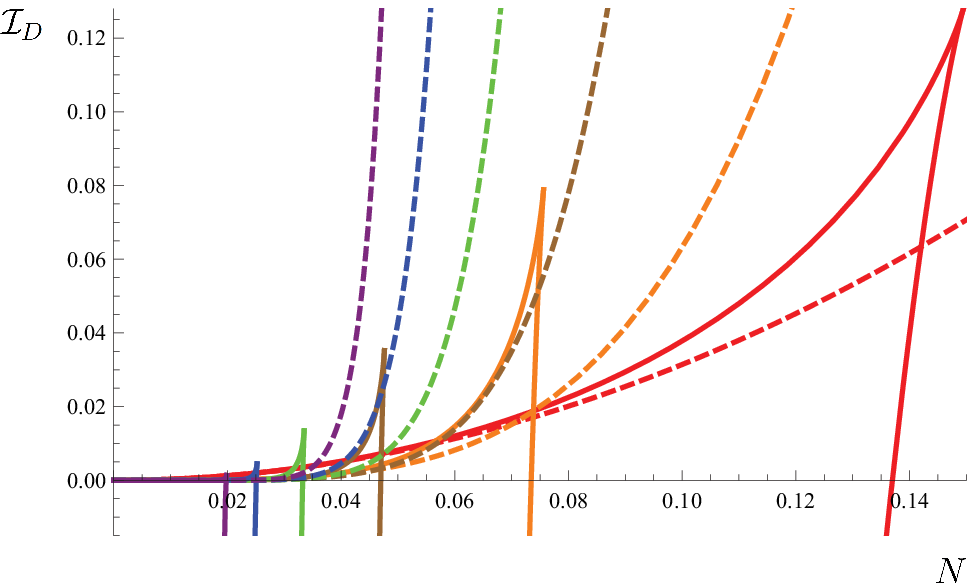}}~~~{\includegraphics[width=8cm]{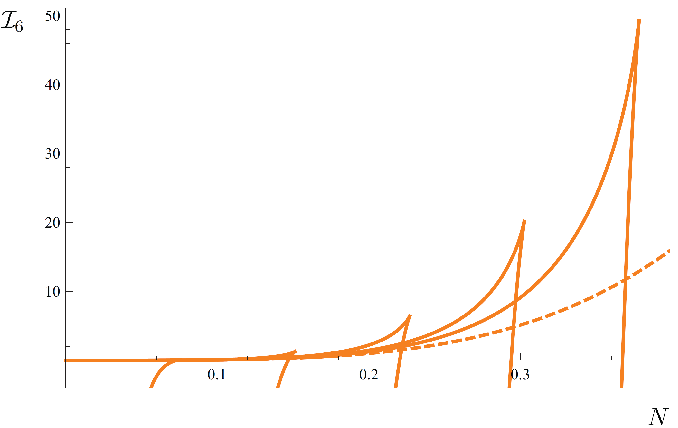}}
\end{center}
~~~~~~~~~~~~~~~~~~~~~~~~~~~~~~~(a)~~~~~~~~~~~~~~~~~~~~~~~~~~~~~~~~~~~~~~~~~~~~~~~~~~~~~~~(b)
\caption{{\footnotesize (a) Plot of the action difference as a function of $N$ for $u$
(from left to right) 1 (the dimension of spacetime $2u+2=4$) to 6  (the dimension of spacetime $2u+2=14$)
for cosmological parameter $l=1$ and pressure $p=3$.
(b) Plot of the six-dimensional action difference ${\cal I}_6$ as a function of $N$
for cosmological parameter $l$ (from left to right) 1 to 5 for pressure $p=3$.}}
\label{figI}
\end{figure}
As shown in Fig. 2, taking the cosmological parameter $l$ and the pressure $p$ as fixed parameters,
the action difference ${\cal I}_{D}$ becomes negative as $N$ increases.
This means that there is the first order phase transition from Taub-NUT-AdS
to Taub-Bolt-AdS at a critical NUT charge.
Note that there is the transition to Taub-Bolt-AdS with $r_{\rm B, +}$ only
since the right vertical solid curves come from the larger branch of the two branches of bolts
where $r_{\rm B, +}$ represents large bolt radii in (\ref{rB}).
As shown in Fig 2. (a), the curves of ${\cal I}_{D}$ move more to the left
as the dimension of spacetime increases.
In Fig 2. (b), it is shown that the curves of ${\cal I}_{6}$ move more to the right as the pressure decreases.
The pressure $p$ becomes zero as the cosmological parameter $l$ goes to infinity
since $p$ is inversely proportional to $l$ i.e., $p={u(2u+1)}/{(8\pi l^2)}$.
In such limit, curves of the action difference
as a function of $N$ correspond to each dashed line in Fig. 2.\\

\begin{figure}[!htbp]
\begin{center}
{\includegraphics[width=12cm]{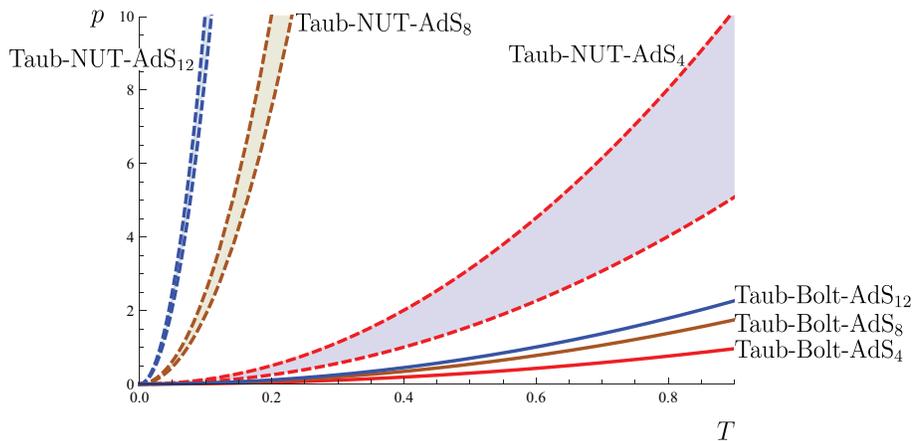}}
\end{center}
\caption{{\footnotesize For the NUT solution, plot of thermodynamically stable range of $p$ as a function of $T$
for given dimension of spacetime (two red dashed curves for $u=1$, two brown dashed curves for $u=3$,
and two blue dashed curve for $u=5$, respectively) and for the Bolt solution,
plot of $p$ as a function of $T$ for given dimension of spacetime
(red solid curve for $u=1$, brown solid curve for $u=3$, and blue solid curve for $u=5$, respectively).}}
\label{figI}
\end{figure}
Requiring both the entropy and the specific heat are positive,
we find the following thermodynamically stable range of $p$ for the NUT solution
\bear\label{Nstable}
\frac{\pi u(2u-1)(u+1)}{T^2}< p <\frac{\pi(2u-1)(u+1)^2}{T^2}.
\eear
Note that there is thermodynamically stable range of $p$ for all odd $u$ only,
since both the entropy and specific heat at a given temperature are not positive for all even $u$
(see e.g., Fig 5. (b)). It is shown in Fig. 3 that such stable range (\ref{Nstable})
is given as the shaded regions between the two curves on the left.

After obtaining $N$ with a change of sign in the action difference ${\cal I}_{D}$ (\ref{diffaction})
through solving ${\cal I}_{D}=0$ for $N$, and substituting into the inverse of the temperature (\ref{inverT})
we may deduce an expression for the critical temperature $T_{c}$
\bear\label{Tc}
T_{c} = \frac{\sqrt{p}}{\sqrt{2\pi}\alpha(u+1)\sqrt{u(2u+1)}},
\eear
by employing the pressure (\ref{pressure}).
Here at $T_{c}$ a phase transition from Taub-NUT-AdS to Taub-Bolt-AdS occurs and
$\alpha$ is constant. For example, for $u=1$, the transition temperature $T_{c}$ reduces
to the result in \cite{Johnson:2014pwa}
\bear
\alpha=\frac{\sqrt{5}-\sqrt{2}}{6},
\eear
and when $u>1$ (beyond four-dimensional spacetime), $\alpha$ is numerically written in Table 3 in Appendix A
since it is difficult to solve the equation with the analytic method for a complicated high-order
polynomial equation ${\cal I}_{D} =0$.
It is shown in Fig. 3 that the rightmost solid curves
(red solid curve for $u=1$, brown solid curve for $u=3$, and blue solid curve for $u=5$, respectively)
denote the lines of transition between the Taub-NUT-AdS and Taub-Bolt-AdS phases
for a given dimension of spacetime and this transition occurs at $T_{c}$ (\ref{Tc}).

\begin{figure}[!htbp]
\begin{center}
{\includegraphics[width=8cm]{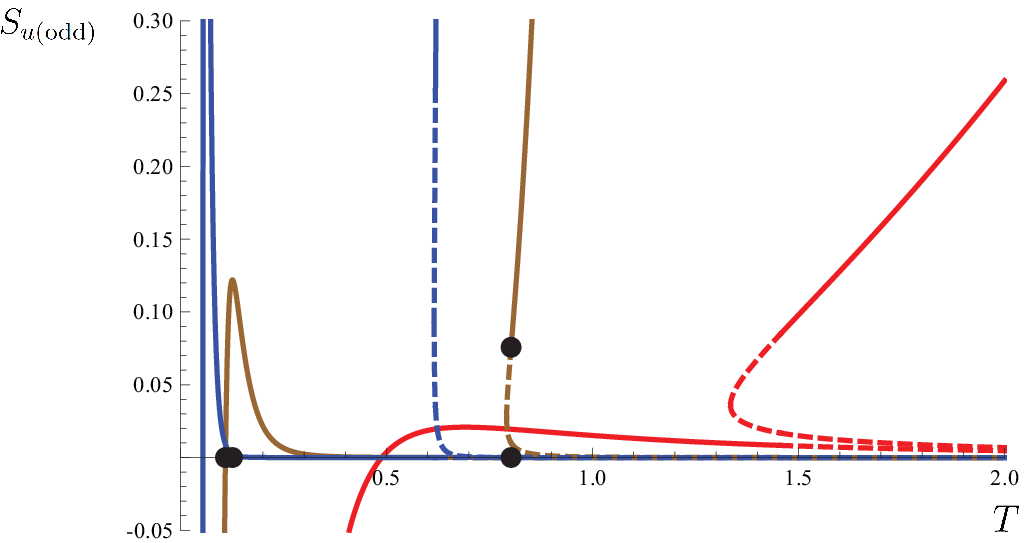}}~~~{\includegraphics[width=8cm]{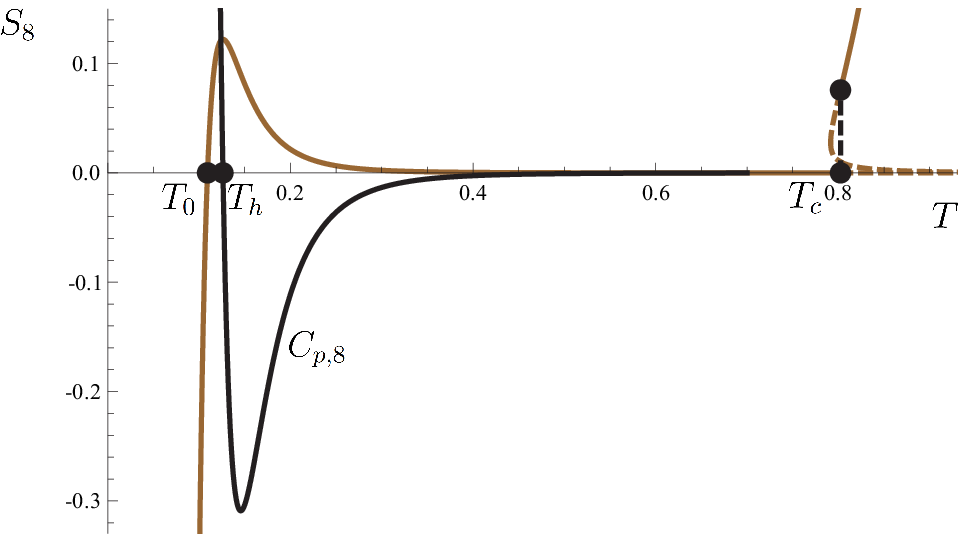}}
\end{center}
~~~~~~~~~~~~~~~~~~~~~~~~~~~~~~~(a)~~~~~~~~~~~~~~~~~~~~~~~~~~~~~~~~~~~~~~~~~~~~~~~~~~~~~~~(b)
\caption{{\footnotesize (a) Plot of the $(2u+2)$-dimensional entropy $S_{u(odd)}$ as a function of temperature $T$
for odd $u$ (red solid curve for $u=1$, brown solid curve for $u=3$, and blue solid curve for $u=5$, respectively)
for pressure $p=3$. (b) Plot the eight-dimensional entropy $S_8$ (brown solid curve)
and specific heat $C_8$ (black solid curve)
as a function of temperature $T$ for pressure $p=3$.}}
\begin{center}
{\includegraphics[width=8cm]{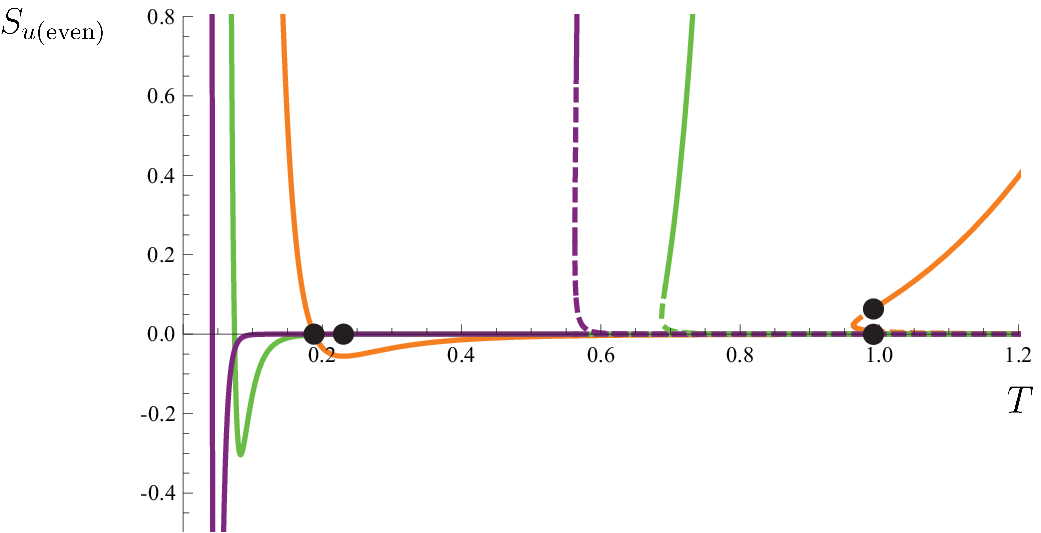}}~~~{\includegraphics[width=8cm]{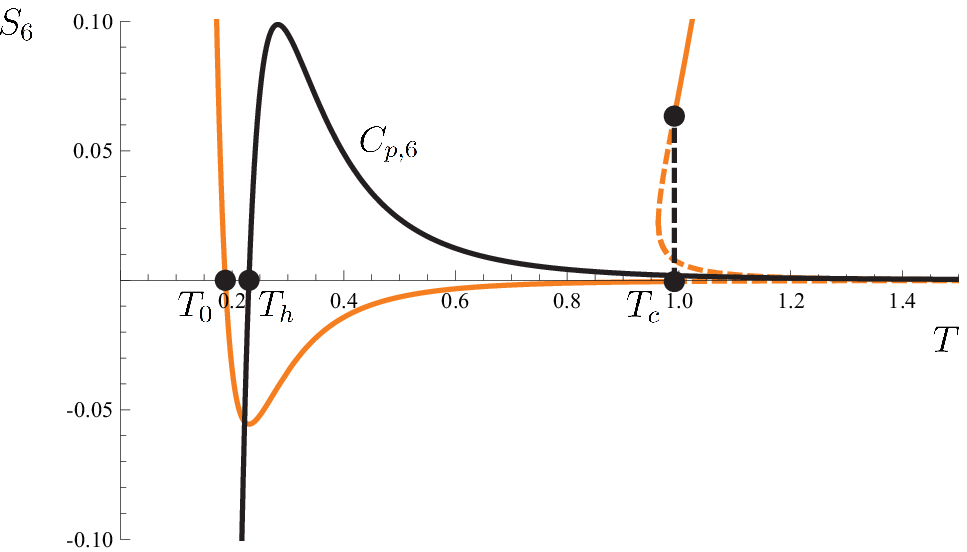}}
\end{center}
~~~~~~~~~~~~~~~~~~~~~~~~~~~~~~~(a)~~~~~~~~~~~~~~~~~~~~~~~~~~~~~~~~~~~~~~~~~~~~~~~~~~~~~~~(b)
\caption{{\footnotesize (a) Plot of the $(2u+2)$-dimensional entropy $S_{u(even)}$ as a function of temperature $T$
for even $u$ (orange solid curve for $u=2$, green solid curve for $u=4$, and purple solid curve for $u=6$, respectively)
for pressure $p=3$. (b) Plot of the six-dimensional entropy $S_6$ (orange solid curve)
and specific heat $C_6$ (black solid curve) as a function of temperature $T$ for pressure $p=3$.}}
\label{figI}
\end{figure}

In Fig 4.and Fig 5., dashed lines represent solutions for higher actions.
As shown in Fig 4. (b) and Fig 5. (b),
the entropy is zero at $T_0$
\bear
T_0 =\frac{\sqrt{p}}{4\sqrt{5\pi}}~~~(u=3)~{\rm and}
~T_0=\frac{\sqrt{p}}{3\sqrt{3\pi}}~~~(u=2),
\eear
and the specific heat is zero at $T_{h}$
\bear
T_{h} =\frac{\sqrt{p}}{2\sqrt{15\pi}}~~~(u=3)~{\rm and}
~T_h=\frac{\sqrt{p}}{3\sqrt{2\pi}}~~~(u=2),
\eear
where at $T_{h}$ the entropy has maximum value (odd $u$) or minimum (even $u$)
(for the generalized form of $T_0$ and $T_h$, see Table 3 in Appendix A).
Any NUT solution in AdS space is thermodynamically stable for all odd $u$
since both the entropy $S_{u(odd)}$ and the specific heat $C_{p,u(odd)}$ become positive between $T_0$ and $T_h$
where $S_{u(odd)}$ and $C_{p,u(odd)}$ denote the $(2u+2)$-dimensional entropy for odd $u$ and
the $(2u+2)$-dimensional specific heat for odd $u$ at constant pressure
(see e.g., Fig. 4 (b)  for $u=3$, and \cite{Johnson:2014pwa} for $u=1$).
It is also found that the thermodynamic nature of the Taub-NUT/Bolt-AdS system dramatically changes
due to odd $u$ or even $u$, i.e., there is stable Bolt solution only for all even $u$.
Thus, here focusing on all odd $u$, we discuss this phase transition to Taub-Bolt-AdS in detail.
In Fig 4. (b), the brown solid curve on the left is the eight-dimensional entropy of the NUT solution,
the black solid curve is the eight-dimensional specific heat of the NUT solution, and
the brown solid curve on the right is
the eight-dimensional entropy of the Bolt solution with the large bolt radii $r_{\rm B,+}$.
Thus possible profiles of the entropy are classified by the value of the temperature $T$:
(i) when $T<T_0$ and for the NUT solution, $S_8$ is negative but $C_{p,8}$ is positive as shown in Fig 4. (b).
(ii) when $T_0<T<T_h$ and for the NUT solution, both $S_8$ and $C_{p,8}$ are positive as shown in Fig 4. (b).
(iii) when $T_h<T<T_c$ and for the NUT solution, $S_8$ is positive but $C_{p,8}$ is negative as shown in Fig 4. (b).
(iv) when $T>T_c$ and for the Bolt solution, $S_8$ is positive as shown in Fig 4. (b)., and $C_{p,8}$ is positive.
Here we do not draw the plot of the specific heat as the function of temperature $T$
since the specific heat $C_{p,8}$ is always positive in the case of Bolt solution with the $r_{\rm B,+}$.
As pointed out in \cite{Johnson:2014yja,Johnson:2014pwa},
in the extended thermodynamics the profiles above may be interpreted as follows:
For (i), the negative entropy comes from some net heat outflow during the creation of the Taub-NUT black hole.
For (ii), the stable Taub-NUT black hole occurs and,
in particular,
since the thermodynamic volume $\scriptsize{-\frac{1}{4\pi}\left(\frac{\sqrt{2u-1}}{2\sqrt{p}}\right)^{2u+1}
\Gamma(-\frac{1}{2}-u)\Gamma(u+1)}$ at $T_0$ and
the thermodynamic volume $\scriptsize{-\frac{1}{4\pi}\left(\frac{\sqrt{u(2u^2+u-1)}}{2(u+1)\sqrt{p}}\right)^{2u+1}
\Gamma(-\frac{1}{2}-u)\Gamma(u+1)}$ at $T_{h}$,
the thermal stable range of the thermodynamic volume is given as
\bear
-\frac{1}{24\sqrt{\pi}p^{\frac{3}{2}}}<V_{\rm NUT,4}<
-\frac{1}{48\sqrt{2\pi}p^{\frac{3}{2}}}~~~({\rm for}~u=1),
\eear
and
\bear
-\frac{25\sqrt{5}}{112\sqrt{\pi}p^{\frac{7}{2}}}<V_{\rm NUT,8}
<-\frac{675\sqrt{15}}{14336\sqrt{\pi}p^{\frac{7}{2}}}~~~({\rm for}~u=3),
\eear
where it can be seen that the thermal stable range of the thermodynamic volume still increases with increasing dimensionality.
For (iii), the Taub-NUT black hole develops an instability
due to the negative specific heat. For (iv), the stable Taub-Bolt black hole finally occurs.

\section{Conclusion}
In the context of the extended thermodynamics,
we considered the higher dimensional Tabu-NUT/Bolt-AdS solution
and obtained their thermal quantities such as the enthalpy, the entropy, the specific heat,
and the thermodynamic volume and so on.

We investigated their phase structure through their thermal
quantities, and found out that the thermodynamic nature of the
Taub-NUT/Bolt-AdS system dramatically changes due to odd or even $u$.
In particular, it was found out that there existed the first order
phase transition from the Taub-NUT-AdS to the Taub-Bolt-AdS for all
odd $u$.

Finally, we also explored a proportional behavior of their thermodynamic quantities with increasing dimensionality
and found that the maximum magnitude of the NUT charge, the magnitude of radii of
small bolt, and the thermal stable range of a thermodynamic volume grow up.

\appendix{Thermodynamic quantities for NUT/Bolt solution}
\begin{center}
\begin{table}[!hbp]
\renewcommand{\arraystretch}{1.4}
\begin{tabular}{|c | c | c| c|} \hline
$D$& $H_{\rm NUT}$ & $V_{\rm NUT}$ & $U_{\rm NUT}$\\ \hline
4  & $N(1-\frac{32N^2\pi}{3}p)$                    & $-\frac{8N^3\pi}{3}$ & $N(1-\frac{3N^2}{l^2})$\\ \hline
6  & $-\frac{32}{3}N^3\pi(1-\frac{24N^2\pi}{5}p)$  & $\frac{128N^5\pi^2}{15}$ & $-\frac{32}{3}N^3(1-\frac{5N^2}{l^2})\pi$\\ \hline
8  & $\frac{384}{5}N^5\pi^2(1-\frac{64N^2\pi}{21}p)$& $-\frac{1024N^7\pi^3}{35}$ & $\frac{384}{5}N^5(1-\frac{7N^2}{l^2})\pi^2$\\ \hline
10 & $-\frac{16384}{35}N^7\pi^3(1-\frac{20N^2\pi}{9}p)$& $\frac{32768N^9\pi^4}{315}$ & $-\frac{16384}{35}N^7(1-\frac{9N^2}{l^2})\pi^3$\\ \hline
n  &$\begin{array}{c}
u(4\pi)^{u-1}\bigg\{\frac{N^{2u-1}\Gamma\left(\frac{3}{2}-u\right)\Gamma\left(u+1\right)}{\sqrt{\pi}(2u-1)}\\
-\frac{16\sqrt{\pi}(u+1)N^{2u+1}\Gamma\left(\frac{3}{2}-u\right)\Gamma\left(u+1\right)}{u(2u-1)(2u+1)}p\bigg\}
\end{array}$
   &$-\frac{u(4\pi)^uN^{2u+1}}{2\sqrt{\pi}}\Gamma(-\frac{1}{2}-u)\Gamma(u)$
   &$\begin{array}{c}\frac{u(2u+1)}{8\pi}\bigg\{(2\sqrt{\pi}N)^{2u-1}\\
-\frac{(2u+1)(2\sqrt{\pi}N)^{2u-1}N^2}{l^2}\bigg\}\\
\times\Gamma(-\frac{1}{2}-u)\Gamma(u+1)
\end{array}$ \\\hline
\end{tabular}
\end{table}
\begin{center}{
Table 1: Summary of thermodynamic quantities for Taub-NUT-AdS}
\end{center}
\end{center}
\vspace{2mm}
\begin{center}
\begin{table}[!hbp]
\renewcommand{\arraystretch}{1.4}
\begin{tabular}{|c | c | c| c|} \hline
$D$& $H_{\rm Bolt}$ & $V_{\rm Bolt}$ & $U_{\rm Bolt}$\\ \hline
4  & \scriptsize{$\frac{r_{\rm B}^2+N^2}{2r_{\rm B}}+\frac{4\pi(r_{\rm B}^4-6N^2r_{\rm B}^2-3N^4)}{3r_{\rm B}}p$}
   & \scriptsize{$\frac{4\pi r_{\rm B}}{3}(r_{\rm B}^2-3N^2)$}
   & \scriptsize{$\frac{r_{\rm B}^2+N^2}{2r_{\rm B}}(1-\frac{3N^2}{l^2})$}\\ \hline
6  &\scriptsize{$\begin{array}{c}\frac{4\pi(r_{\rm B}^4-6N^2r_{\rm B}^2-3N^4)}{3r_{\rm B}}\\
+\frac{16\pi^2(r_{\rm B}^6-5N^2r_{\rm B}^4+15N^4r_{\rm B}^2+5N^6)}{5r_{\rm B}}p
\end{array}$}
   & \scriptsize{$\frac{16\pi^2 r_{\rm B}}{15}(3r_{\rm B}^4-10N^2r_{\rm B}^2+15N^4)$}
   & \scriptsize{$\frac{4\pi(r_{\rm B}^4-6N^2r_{\rm B}^2-3N^4)}{3r_{\rm B}}(1-\frac{5N^2}{l^2})$}\\ \hline
8  & \scriptsize{$\begin{array}{c}\frac{24\pi^2(r_{\rm B}^6
-5N^2r_{\rm B}^4+15N^4r_{\rm B}^2+5N^6)}{5r_{\rm B}}\\
+\frac{64\pi^3}{35r_{\rm B}}(5r_{\rm B}^8-28N^2r_{\rm B}^6\\
+70N^4r_{\rm B}^4-140N^6r_{\rm B}^2-35N^8)p
\end{array}$}
   &\scriptsize{$\begin{array}{c}\frac{64\pi^3 r_{\rm B}}{35}(5r_{\rm B}^6-21N^2r_{\rm B}^4\\
   +35N^4r_{\rm B}^2-35N^6)
\end{array}$}
   & \scriptsize{$\begin{array}{c}\frac{24\pi^2}{5r_{\rm B}}(r_{\rm B}^6-5N^2r_{\rm B}^4\\
   +15N^4r_{\rm B}^2+5N^6)\\
   \times(1-\frac{7N^2}{l^2})
\end{array}$}\\ \hline
10 &  \scriptsize{$\begin{array}{c}\frac{128\pi^3}{35r_{\rm B}}
(5r_{\rm B}^8-28N^2r_{\rm B}^6\\
+70N^4r_{\rm B}^4-140N^6r_{\rm B}^2-35N^8)\\
+\frac{256\pi^4}{63r_{\rm B}}(7r_{\rm B}^{10}-45N^2r_{\rm B}^8+126N^4r_{\rm B}^6\\
-210N^6r_{\rm B}^4+315N^8r_{\rm B}^2+63N^{10})p
\end{array}$}
   & \scriptsize{$\begin{array}{c}\frac{256\pi^4 r_{\rm B}}{315}(35r_{\rm B}^8-180N^2r_{\rm B}^6\\
   +378N^4r_{\rm B}^4-420N^6r_{\rm B}^2+315N^8)
\end{array}$}
   & \scriptsize{$\begin{array}{c}\frac{128\pi^3}{35r_{\rm B}}(5r_{\rm B}^8-28N^2r_{\rm B}^6\\
   +70N^4r_{\rm B}^4+40N^6r_{\rm B}^2\\
   -35N^8)(1-\frac{9N^2}{l^2})
\end{array}$}\\ \hline
n  &$\scriptsize{\begin{array}{c}
\frac{u(4 \pi)^{u-1}}{2}
\bigg\{{\sum_{i=0}^{u}
\left( { u \atop i} \right)}
\frac{(-1)^i N^{2i}r_{\rm B}^{2u-2i-1}}{(2u-2i-1)}\\
+\frac{8\pi}{u}\sum_{i=0}^{u+1}
\left( { u+1 \atop i} \right)
\frac{(-1)^iN^{2i}r_{\rm B}^{2u-2i+1}}{(2u-2i+1)}p\bigg\}
\end{array}}$
   &$\scriptsize{\begin{array}{c}
   \frac{\pi^{u}(2r_{\rm B})^{u-1}}{2u+1}
   \bigg\{-2(N^2-r_{\rm B}^2)
(1-\frac{N^2}{r_{\rm B}^2})^u\\
+r_{\rm B}^2(\frac{N}{r_{\rm B}})^{2u+1}
{\rm B} (\frac{N^2}{r_{\rm B}^2},\frac{1}{2}-u,u+1)\bigg\}
\end{array}}$
   &${\scriptsize\begin{array}{c}
\frac{(4\pi)^{u-1}}{4l^2}uN^{2u-1}\\
\times\{(2u+1)N^2-l^2\}\\
\times
{{\rm B} (\frac{N^2}{r_{\rm B}^2},\frac{1}{2}-u,u+1)}
\end{array}}$ \\\hline
\end{tabular}
\end{table}
\begin{center}{
Table 2: Summary of thermodynamic quantities for Taub-Bolt-AdS}
\end{center}
\end{center}
\newpage
\begin{center}
\begin{table}[!hbp]
\renewcommand{\arraystretch}{1.4}
\begin{tabular}{|c | c | c| c| c|} \hline
$D$& ${\cal I}_D$ & $T_0$ & $T_{h}$ & $T_{c}$ \\ \hline
4  & \scriptsize{$\begin{array}{c}
-\frac{2N\pi(r_{\rm B}-N)^2}{r_{\rm B}l^2}(r_{\rm B}^2+2Nr_{\rm B}+3N^2-l^2)
\end{array}$}                    & $\frac{\sqrt{p}}{2\sqrt{\pi}}$ & $\frac{\sqrt{p}}{\sqrt{2\pi}}$&$0.84077\sqrt{p}$ \\ \hline
6  &\scriptsize{$\begin{array}{c}
\frac{4N\pi^2(r_{\rm B}-N)^3}{r_{\rm B}l^2}\bigg\{3r_{\rm B}^3+9Nr_{\rm B}^2+13N^2r_{\rm B}+15N^3\\
-(r_{\rm B}+3N)l^2\bigg\}
\end{array}$}   & $\frac{\sqrt{p}}{3\sqrt{3\pi}}$& $\frac{\sqrt{p}}{3\sqrt{2\pi}}$  &$0.57253\sqrt{p}$ \\ \hline
8  & \scriptsize{$\begin{array}{c}
-\frac{64N\pi^3(r_{\rm B}-N)^4}{5r_{\rm B}l^2}\bigg\{5r_{\rm B}^4+20Nr_{\rm B}^3+36N^2r_{\rm B}^2\\
+44N^3r_{\rm B}+35N^4\\
-(r_{\rm B}^2+4Nr_{\rm B}+5N^2)l^2\bigg\}
\end{array}$} & $\frac{\sqrt{p}}{4\sqrt{5\pi}}$& $\frac{\sqrt{p}}{2\sqrt{15\pi}}$ &$0.46324\sqrt{p}$ \\ \hline
10 & \scriptsize{$\begin{array}{c}
\frac{64N\pi^4(r_{\rm B}-N)^5}{7r_{\rm B}l^2}\bigg\{35r_{\rm B}^5+175Nr_{\rm B}^4+390N^2r_{\rm B}^3\\
+550N^3r_{\rm B}^2+55N^43r_{\rm B}+315N^5\\
-(5r_{\rm B}^3+25Nr_{\rm B}^2+47N^23r_{\rm B}+35N^3)l^2\bigg\}
\end{array}$}& $\frac{\sqrt{p}}{5\sqrt{7\pi}}$&$\frac{\sqrt{p}}{2\sqrt{35\pi}}$ &$0.40000\sqrt{p}$\\ \hline
n  & $\scriptsize{\begin{array}{c}
\frac{(4\pi)^u}{8r_{\rm B}l^2}\bigg[2N(N^2-r_{\rm B}^2)
(u+1)(1-\frac{N^2}{r_{\rm B}^2})^u\\
+(2uN^2-l^2)\bigg\{\\
(u+1)N^{2u}{\rm B}(\frac{N^2}{r_{\rm B}^2},\frac{1}{2}-u,u+1)\\
-\frac{2N^{2u}}{\sqrt{\pi}}\Gamma(\frac{1}{2}-u)\Gamma(2+u)\bigg\}\bigg]
\end{array}}$
   & $\frac{\sqrt{p}}{\sqrt{\pi}(u+1)\sqrt{2u-1}}$
   & $\frac{\sqrt{p}}{\sqrt{\pi}\sqrt{u(2u^2+u-1)}}$
   & $\frac{\sqrt{p}}{\sqrt{2\pi}\alpha(u+1)\sqrt{u(2u+1)}}$ \\\hline
\end{tabular}
\end{table}
\begin{center}{
Table 3: Summary of thermodynamic quantities for exploring the phase structure}
\end{center}
\end{center}



\begin{thebibliography}{100}
\bibitem{Bekenstein:1973ur}
  J.~D.~Bekenstein,
  Phys.\ Rev.\ D {\bf 7}, 2333 (1973).

\bibitem{Bardeen:1973gs}
  J.~M.~Bardeen, B.~Carter and S.~W.~Hawking,
  Commun.\ Math.\ Phys.\  {\bf 31}, 161 (1973).

\bibitem{Hawking:1982dh}
  S.~W.~Hawking and D.~N.~Page,
  Commun.\ Math.\ Phys.\  {\bf 87}, 577 (1983).



\bibitem{Lousto:1994jd}
  C.~O.~Lousto,
  Phys.\ Rev.\ D {\bf 51}, 1733 (1995)
  [gr-qc/9405048].

\bibitem{Peca:1998cs}
  C.~S.~Peca and J.~Lemos, P.S.,
  Phys.\ Rev.\ D {\bf 59}, 124007 (1999)
  [gr-qc/9805004].

\bibitem{Banerjee:2010bx}
  R.~Banerjee, S.~K.~Modak and S.~Samanta,
  Phys.\ Rev.\ D {\bf 84}, 064024 (2011)
  [arXiv:1005.4832 [hep-th]].

\bibitem{Myung:2012xc}
  Y.~S.~Myung,
  Eur.\ Phys.\ J.\ C {\bf 72}, 2116 (2012)
  [arXiv:1203.1367 [hep-th]].



\bibitem{Chamblin:1999tk}
  A.~Chamblin, R.~Emparan, C.~V.~Johnson and R.~C.~Myers,
  Phys.\ Rev.\ D {\bf 60}, 064018 (1999)
  [hep-th/9902170].

\bibitem{Cvetic:1999ne}
  M.~Cvetic and S.~S.~Gubser,
  JHEP {\bf 9904}, 024 (1999)
  [hep-th/9902195].

\bibitem{Caldarelli:1999xj}
  M.~M.~Caldarelli, G.~Cognola and D.~Klemm,
  Class.\ Quant.\ Grav.\  {\bf 17}, 399 (2000)
  [hep-th/9908022].

\bibitem{Gubser:2000mm}
  S.~S.~Gubser and I.~Mitra,
  JHEP {\bf 0108}, 018 (2001)
  [hep-th/0011127].

\bibitem{Cai:2001dz}
  R.~-G.~Cai,
  Phys.\ Rev.\ D {\bf 65}, 084014 (2002)
  [hep-th/0109133].

\bibitem{Dey:2004yt}
  T.~K.~.Dey,
  Phys.\ Lett.\ B {\bf 595}, 484 (2004)
  [hep-th/0406169].

\bibitem{Dey:2006ds}
  T.~K.~Dey, S.~Mukherji, S.~Mukhopadhyay and S.~Sarkar,
  JHEP {\bf 0704}, 014 (2007)
  [hep-th/0609038].

\bibitem{Banerjee:2007by}
  N.~Banerjee and S.~Dutta,
  JHEP {\bf 0707}, 047 (2007)
  [arXiv:0705.2682 [hep-th]].

\bibitem{Dey:2007vt}
  T.~K.~Dey, S.~Mukherji, S.~Mukhopadhyay and S.~Sarkar,
  JHEP {\bf 0709}, 026 (2007)
  [arXiv:0706.3996 [hep-th]].

\bibitem{Konoplya:2008rq}
  R.~A.~Konoplya and A.~Zhidenko,
  Phys.\ Rev.\ D {\bf 78}, 104017 (2008)
  [arXiv:0809.2048 [hep-th]].

\bibitem{Banerjee:2011au}
  R.~Banerjee and D.~Roychowdhury,
  JHEP {\bf 1111}, 004 (2011)
  [arXiv:1109.2433 [gr-qc]].


\bibitem{Kastor:2009wy}
  D.~Kastor, S.~Ray and J.~Traschen,
  Class.\ Quant.\ Grav.\  {\bf 26} (2009) 195011
  [arXiv:0904.2765 [hep-th]].

\bibitem{Cvetic:2010jb}
  M.~Cvetic, G.~W.~Gibbons, D.~Kubiznak and C.~N.~Pope,
  Phys.\ Rev.\ D {\bf 84}, 024037 (2011)
  [arXiv:1012.2888 [hep-th]].

\bibitem{Dolan:2011xt}
  B.~P.~Dolan,
  Class.\ Quant.\ Grav.\  {\bf 28}, 235017 (2011)
  [arXiv:1106.6260 [gr-qc]].

\bibitem{Kubiznak:2012wp}
  D.~Kubiznak and R.~B.~Mann,
  JHEP {\bf 1207}, 033 (2012)
  [arXiv:1205.0559 [hep-th]].

\bibitem{Gunasekaran:2012dq}
  S.~Gunasekaran, R.~B.~Mann and D.~Kubiznak,
  JHEP {\bf 1211}, 110 (2012)
  [arXiv:1208.6251 [hep-th]].

\bibitem{Hendi:2012um}
  S.~H.~Hendi and M.~H.~Vahidinia,
  Phys.\ Rev.\ D {\bf 88}, no. 8, 084045 (2013)
  [arXiv:1212.6128 [hep-th]].

\bibitem{Cai:2013qga}
  R.~-G.~Cai, L.~-M.~Cao, L.~Li and R.~-Q.~Yang,
  JHEP {\bf 1309}, 005 (2013)
  [arXiv:1306.6233 [gr-qc]].

\bibitem{Xu:2013zea}
  W.~Xu, H.~Xu and L.~Zhao,
  arXiv:1311.3053 [gr-qc].

\bibitem{Zou:2013owa}
  D.~C.~Zou, S.~J.~Zhang and B.~Wang,
  Phys.\ Rev.\ D {\bf 89}, 044002 (2014)
  [arXiv:1311.7299 [hep-th]].

\bibitem{Kubiznak:2014zwa}
  D.~Kubiznak and R.~B.~Mann,
  arXiv:1404.2126 [gr-qc].

\bibitem{Zou:2014mha}
  D.~C.~Zou, Y.~Liu and B.~Wang,
  arXiv:1404.5194 [hep-th].

\bibitem{Johnson:2014yja}
  C.~V.~Johnson,
  arXiv:1404.5982 [hep-th].

\bibitem{Liu:2014gvf}
  Y.~Liu, D.~C.~Zou and B.~Wang,
  arXiv:1405.2644 [hep-th].

\bibitem{Xu:2014tja}
  H.~Xu, W.~Xu and L.~Zhao,
  arXiv:1405.4143 [gr-qc].

\bibitem{Ma:2014vxa}
  M.~-S.~Ma and Y.~-Q.~Ma,
  arXiv:1405.7609 [hep-th].

\bibitem{Xu:2014kwa}
  W.~Xu and L.~Zhao,
  arXiv:1405.7665 [gr-qc].

\bibitem{MacDonald:2014zaa}
  S.~MacDonald,
  arXiv:1406.1257 [hep-th].

\bibitem{Johnson:2014pwa}
  C.~V.~Johnson,
  arXiv:1406.4533 [hep-th].

\bibitem{Frassino:2014pha}
  A.~M.~Frassino, D.~Kubiznak, R.~B.~Mann and F.~Simovic,
  arXiv:1406.7015 [hep-th].

\bibitem{Dolan:2014vba}
  B.~P.~Dolan, A.~Kostouki, D.~Kubiznak and R.~B.~Mann,
  arXiv:1407.4783 [hep-th].




\bibitem{Chamblin:1998pz}
  A.~Chamblin, R.~Emparan, C.~V.~Johnson and R.~C.~Myers,
  Phys.\ Rev.\  D {\bf 59}, 064010 (1999)
  [arXiv:hep-th/9808177].


\bibitem{Hawking:1998ct}
  S.~W.~Hawking, C.~J.~Hunter and D.~N.~Page,
  Phys.\ Rev.\  D {\bf 59}, 044033 (1999)
  [arXiv:hep-th/9809035].


\bibitem{Mann:1999bt}
  R.~B.~Mann,
  Phys.\ Rev.\  D {\bf 61}, 084013 (2000)
  [arXiv:hep-th/9904148].

\bibitem{Awad:2000gg}
  A.~Awad and A.~Chamblin,
  Class.\ Quant.\ Grav.\  {\bf 19}, 2051 (2002)
  [arXiv:hep-th/0012240].


\bibitem{Clarkson:2002uj}
  R.~Clarkson, L.~Fatibene and R.~B.~Mann,
  Nucl.\ Phys.\  B {\bf 652}, 348 (2003)
  [arXiv:hep-th/0210280].


\bibitem{Astefanesei:2004ji}
  D.~Astefanesei, R.~B.~Mann and E.~Radu,
  Phys.\ Lett.\  B {\bf 620}, 1 (2005)
  [arXiv:hep-th/0406050].


\bibitem{Astefanesei:2004kn}
  D.~Astefanesei, R.~B.~Mann and E.~Radu,
  JHEP {\bf 0501}, 049 (2005)
  [arXiv:hep-th/0407110].


\bibitem{Mann:2003zh}
  R.~B.~Mann and C.~Stelea,
  Class.\ Quant.\ Grav.\  {\bf 21}, 2937 (2004)
  [arXiv:hep-th/0312285].


\bibitem{Mann:2005ra}
  R.~B.~Mann and C.~Stelea,
  Phys.\ Lett.\  B {\bf 634}, 448 (2006)
  [arXiv:hep-th/0508203].



\end{thebibliography}
\end{document}